\documentclass{article}
\usepackage[utf8]{inputenc}
\usepackage{amsmath}
\pdfoutput=1

\begin{document}

\begin{center}
  
\textbf{Eliminating black holes singularity in Brane world Cosmology.}\\
\vspace{5mm}
Poula Tadros$^1$ and Mohamed Assaad Abdel-Raouf$^2$\\
\vspace{5mm}
1 Department of Physics and Astronomy, University of Turku, 20500, Turku,Finland.\\
2 Physics Department, Faculty of Science, Ain Shame University, Cairo, Egypt.\\

Emails: pttadr@utu.fi , Assaad@sci.asu.edu.eg
\end{center}

\vspace{10mm}
\begin{abstract}
    In recent works, a construction was proposed resulting emergent universes inside black holes. This result can be obtained from a 4D black hole embedded in a 5D spacetime with the fifth dimension compactified on a circle $[0,2]$ (0 and 2 are identified) such that the two branes are at 0 and 1. In the present work, we study this setup by deriving particles' equations of motion in the new universes, based on redefining energy and angular momentum. This leads to disappearance of the singularity in centers of black holes in classical General Relativity.
\end{abstract}


\section{Introduction}
Black holes were introduced as solutions for Einstien's 
field equations by K. Schwarzchild (For English translation see Ref. $[1]$). Schwarzchild solution
 represents a non rotating, non charged spherically symmetric black hole. However, this solution suffers from a singularity at the center of the black hole. In the
 context of string theory it is argued that black holes emerge naturally (Reviewed in Ref. $[2,3]$).\\
Recently, It was suggested that inside the horizon of a black hole there is an emergent cosmology i.e. a new universe $[4]$ using a topological soliton called S-branes $[5]$ to violate the null energy condition so eliminating the singularity.\\

Cosmological perturbations were studied in the emergent universe $[6]$, and a mechanism of reheating $[7]$ was proposed using the decay of the S-brane as an initiator for the big bang and the radiation dominated era. This construction solves the black hole information paradox as the information falling into the black hole emerges in the new universe without any loss.\\

In the present work, we study a particular metric of a 4D black hole embedded in a 5D spacetime within the context of Brane world cosmology. Introducing an additional compact dimension as the fifth dimension we prove that particles falling into the black hole from the visible or the hidden brane do not experience a singularity with the assumption that the cosmological constant is negative or zero inside the horizon. Thus, an emergent cosmology can be defined inside the horizon between the two branes.\\

\section{Mathematical Analysis}
In this setup a 4D black hole in our universe is viewed as embedded in a 5D world. The metric for this black hole is defined as in Ref. [8] to be 
\begin{equation}
ds^2=\frac{\Lambda y^2}{3}(f(r)dt^2-\frac{dr^2}{f(r)}-r^2(d\theta^2+\sin^2(\theta)d\phi^2))-dy^2
\end{equation}
where $(t,r,\theta,\phi)$ are the usual spherical coordinates, y is the coordinate through the fifth dimension(the compact dimension), and $f(r)=1-\frac{2m}{r}-\frac{\Lambda r^2}{3}$ where $\Lambda$ is the cosmological constant.\\

By spherical symmetry of the system on each y=constant plane, we can set $\theta=\frac{\pi}{2}$.

The first geodesic equation (the equation for t) is given by 
\begin{equation}
    \frac{d^2t}{d\tau^2}+\frac{f'(r)}{2f(r)}\frac{dt}{d\tau} \frac{dr}{d\tau}+\frac{1}{y}\frac{dt}{d\tau}\frac{dy}{d\tau}=0,
\end{equation}
where f'(r) is the derivative of f with respect to r.\\
This equation can be integrated to give
\begin{equation}
    \frac{dt}{d\tau}= \frac{E}{y\sqrt{|f|}}
\end{equation}
where E is a constant which represents the total energy of the system in our model.\\

The second geodesic equation(the equation for $\phi$)is given by
\begin{equation}
\frac{d^2\phi}{d\tau^2}+\frac{1}{r}\frac{d\phi}{d\tau}\frac{dr}{d\tau}+\frac{1}{y}\frac{d\phi}{d\tau}\frac{dy}{d\tau}=0,
\end{equation}
Which can be integrated to
\begin{equation}
    \frac{d\phi}{d\tau}=\frac{L}{ry},
\end{equation}
where L is a constant which represents the total angular momentum of the system in our model.\\
The equation for r is given by
\begin{equation}
\frac{d^2r}{d\tau^2}+\frac{f(r)f'(r)}{2}(\frac{dt}{d\tau})^2-\frac{f'(r)}{2f(r)}(\frac{dr}{d\tau})^2+r f(r)(\frac{d\phi}{d\tau})^2+\frac{1}{y}\frac{dr}{d\tau} \frac{dy}{d\tau}=0.
\end{equation}
After using equations (3),(5) and dividing by $E^2$ (recalling that the impact parameter b is given by $b=\frac{L}{E}$), we get
\begin{equation}
    \frac{1}{E^2} \frac{d^2r}{d\tau^2}+\frac{f'(r)}{2y^2}-\frac{1}{2E^2}\frac{dr}{d\tau} \frac{d}{d\tau}(\ln|f(r)|)+\frac{b^2f(r)}{ry^2}+\frac{1}{E^2}\frac{dr}{d\tau} \frac{d}{d\tau}(\ln|y|)=0.
\end{equation}

The equation for y is given by
\begin{equation}
\frac{d^2y}{d\tau^2}+\frac{\Lambda yf(r)}{3}(\frac{dt}{d\tau})^2-\frac{\Lambda y}{3f(r)}(\frac{dr}{d\tau})^2-\frac{\Lambda r^2 y}{3}(\frac{d\phi}{d\tau})^2=0.
\end{equation}

With similar calculations as the previous equation we get
\begin{equation}
    \frac{d^2y}{d\tau^2}-\frac{\Lambda y}{3f(r)}(\frac{dr}{d\tau})^2=\frac{\Lambda E^2}{3y}(b^2-1).
\end{equation}

In the Ekpyrotic scenario where there exist two nearly parallel D-branes one containing our universe(the visible brane), and another brane(the hidden brane). We can redefine y such that the distance between the two branes is 1.

Since particles (except gravitons) are restricted to move on one of the branes, we have two possible values for the impact parameter(b) assuming that the black hole is in the visible brane. If the particle comes from the Hidden brane to the position "above" the black hole, then $b=1$. if the light comes from the visible brane, then $b=0$. Let us examine both cases.\\

\textbf{Case 1: $b=1$}.\\
Equations (7) and (9) will be
\begin{equation}
    \frac{1}{E^2}\frac{d}{d\tau}(\ln(|\frac{dr}{d\tau})|)-\frac{1}{2E^2}\frac{d}{d\tau}(\ln(|f(r)|))+\frac{1}{E^2}\frac{d}{d\tau}(\ln(|y|))+\frac{f}{ry^2}+\frac{f'(r)}{2y^2}=0.
\end{equation}
and
\begin{equation}
    \frac{d^2y}{d\tau^2}=\frac{\Lambda y}{3f(r)}(\frac{dr}{d\tau})^2.
\end{equation}
Using coordinate time instead of the parameter $\tau$ we get
\begin{equation}
    \frac{d}{dr}(\frac{1}{\sqrt{|f(r)|}y}\frac{dy}{dt})=\frac{\Lambda}{3f(r)\sqrt{|f(r)|}}(\frac{dr}{dt})^2.
\end{equation}

\begin{equation}
    \frac{d}{dt}(\ln(|\frac{E}{f(r)}\frac{dr}{dt}|)=\frac{\sqrt{|f(r)|}}{Ey^2}(\frac{f(r)}{r}+\frac{f'(r)}{2}).
\end{equation}
These are integrated to
\begin{equation}
    \frac{dr}{dt}=\frac{f(r)}{E} \exp(\frac{1}{E}\int \frac{\sqrt{|f(r)|}}{y^2}(\frac{f(r)}{r}+\frac{f'(r)}{2})dr )
\end{equation}

\begin{equation}
    y=\exp(\int dt(\sqrt{|f(r)|}\int dt'(\frac{\Lambda}{3f\sqrt{|f(r)|}}(\frac{dr}{dt})^2))
\end{equation}

We examine the limits of y as r tends to zero and infinity.\\
Far before the interaction the velocity of the particle is constant i.e. $\frac{dr}{dt} \rightarrow 0$ and $f \rightarrow - \infty$ so 
\begin{equation}
\lim_{r \to \infty}(y)=1
\end{equation}
which is exactly what is expected as the particle is restricted to be on the brane with y=1 before the interaction.\\

The limit as r tends to zero can be deduced from the fact that the velocity of particles tend to zero on approaching the singularity ${[9]}$ i.e. $\frac{dr}{dt} \rightarrow 0$ and $f \rightarrow -\infty$.
\begin{equation}
\lim_{r \to 0}(y)=1,
\end{equation}
i.e. a particle in the hidden brane will remain in the hidden brane in the discussed limits.\\

\textbf{Case 2: $b=0$}.\\
Equations (7) and (9) give
\begin{equation}
\frac{1}{E^2}\frac{d^2r}{d\tau^2}+\frac{f'(r)}{2y^2}-\frac{1}{2E^2}\frac{dr}{d\tau}\frac{d}{d\tau}(\ln|f(r)|)+\frac{1}{E^2}\frac{dr}{d\tau}\frac{d}{d\tau}(\ln(y))=0
\end{equation}
and
\begin{equation}
\frac{d^2y}{d\tau^2}-\frac{\Lambda y}{3f(r)}(\frac{dr}{d\tau})^2=-\frac{\Lambda E^2}{3y}
\end{equation}

Using coordinate time and with similar calculations as the first case we get
 
 \begin{equation}
     y=\exp(\int dt\sqrt{|f(r)|}[\int dt \frac{\Lambda \sqrt{|f(r)|}}{3} (\frac{1}{E^2}\exp(-\int dr \frac{f(r)f'(r)}{(\frac{dr}{dt})^2})-1)]).
 \end{equation}
Examining the same limits we get
\begin{equation}
    \lim_{r \to \infty}(y)=0,
\end{equation}
because as r approaches infinity $f \rightarrow -\infty$, $f` \rightarrow - \infty$ and $\frac{dr}{dt} \rightarrow 0$, as expected as particles are restricted to be on the visible brane.\\

\begin{equation}
    \lim_{r \to 0}(y)=  \left \{
    \begin{array}{cc} 1 & \quad 
    \Lambda = 0 \\
     0&
    \quad \Lambda < 0  
    \end{array}
    \right .
 \end{equation}

This is because $r \rightarrow 0$ implies $f \rightarrow -\infty$, $f` \rightarrow \infty$. This converges only if $\Lambda \leq 0$ inside the horizon.\\
\section{Discussion}
We have shown that in the case of a non rotating black hole in 4D in 5D spacetime with one compact dimension, the setup of a model where there are two branes on which all particles (except gravitons) are restricted to move, a particle falling radially into the black hole ends up in the bulk between the two branes.\\

This suggests that a new emergent cosmology can be defined in the bulk inside the horizon. The shape of the emergent universe is determined by the solution of eq.(20). The most important feature is that this universe is free of physical singularities. Moreover, particles in the hidden brane never reach the visible brane, that is why we do not observe white holes in our universe in a finite time.\\

In the new universe the particles velocity decreases as they approach a certain point ($r=0$), this can be interpreted as an expanding universe with the center of expansion at $r=0$. From the viewpoint of the hidden brane the black hole looks like a potential well and the emergent universe is trapped inside it. while the visible brane sees the black hole as a bridge to the hidden brane through the bulk. Thus the two viewpoints see a universe in the bulk. Adopting the setup in which the two branes have universe-antiuniverse pair (universe on the visible brane and antiuniverse on the hidden brane), then on any of the two branes one sees a universe, or an antiuniverse. This setup has useful features such as that the matter-antimatter asymmetry is avoided, and the system is CPT symmetric [10].\\

These results agree with Ref. [4] that an emergent universe can be defined inside the horizon from the point of view of a three dimensional observer. However, in this paper we argue that it is in fact in an additional dimension. Thus, we can identify the emergent universe discussed here and the emergent universe in Ref. [4], this can be done by adopting the following timeline: At the beginning of the formation of the black hole the high density deforms the two branes in a certain region until they collide. An S-brane was formed in the interface of the collision, then the S-brane decays inducing reheating as in Ref. [5].\\

\section{References.}

$[1]$ K. Schwarzschild, Sitzungsberichte der Königlich Preuss. Akad. der Wissenschaften $ (1999)$ 189.\\
$[2]$ S. De Haro, J. van Dongen, Visser, M. and J. Butterfield, Stud. Hist. Philos. Sci. Part B - Stud. Hist. Philos. Mod. Phys. $69 (2020)$ 82.\\
$[3]$ J. van Dongen, S. De Haro, M. Visser, and J. Butterfield, Stud. Hist. Philos. Sci. Part B - Stud. Hist. Philos. Mod. Phys. $69  (2020) 112$.\\
$[4]$ R. Brandenberger, L. Heisenberg, and J. Robnik, J. High Energy Phys. 2021 (2021) 90.  \\
$[5]$ M. Gutperle, and A. Strominger, J. High Energy Phys. 6 (2002) 395.\\
$[6]$ R. Brandenberger, and Z. Wang, Phys. Rev. D 102 (2020) 23516.\\
$[7]$ R. Brandenberger, K. Dasgupta, and Z. Wang, Phys. Rev. D 102 (2020) 063514.\\
$[8]$ P. Wesson, H. Mashhoon, and H. Liu, Modern Physics Letters A, 12 (1997) 2309. \\
$[9]$ A. T. Augousti, P. Gusin, B. Kuśmierz, J. Masajada and A. Radosz, Gen. Relativ. Gravit. (2018) 50.\\
$[10]$ L. Boyle,K. Finn, and N. Turok, Phys. Rev. Lett. 121 (2018) 1.\\

\end{document}